\documentclass[aps,prl,reprint,superscriptaddress,twocolumn,floatfix,nofootinbib]{revtex4-2}
\usepackage[utf8]{inputenc}
\usepackage[T1]{fontenc}
\usepackage{amsmath}
\usepackage{amsfonts}
\usepackage{amssymb}
\usepackage[dvipsnames]{xcolor}
\usepackage{graphicx}
\usepackage{physics}
\usepackage{textcomp}
\usepackage{dsfont}
\usepackage{lipsum}
\usepackage{bbold}
\usepackage{tabularx}
\usepackage{longtable}
\usepackage{array}
\usepackage{nicefrac, xfrac}

\emergencystretch=\maxdimen
\allowdisplaybreaks

\begin{document}

\title{Methods for transverse and longitudinal spin-photon coupling in silicon quantum dots with intrinsic spin-orbit effect}

\author{Kevin S. Guo}
\email[Corresponding Author: ]{kevin.s.guo@unsw.edu.au}
\thanks{These authors contributed equally to this work.}
\affiliation{School of Electrical Engineering and Telecommunications, The University of New South Wales, Sydney, NSW 2052, Australia}
\author{MengKe Feng}
\email[Corresponding Author: ]{mengke.feng@unsw.edu.au}
\thanks{These authors contributed equally to this work.}
\affiliation{School of Electrical Engineering and Telecommunications, The University of New South Wales, Sydney, NSW 2052, Australia}
\author{Jonathan Y. Huang}
\affiliation{School of Electrical Engineering and Telecommunications, The University of New South Wales, Sydney, NSW 2052, Australia}
\author{Will Gilbert}
\affiliation{School of Electrical Engineering and Telecommunications, The University of New South Wales, Sydney, NSW 2052, Australia}
\affiliation{Diraq, Sydney, NSW 2052, Australia}
\author{Kohei M. Itoh}
\affiliation{School of Fundamental Science and Technology, Keio University, 3-14-1 Hiyoshi, Kohokuku, Yokohama, 223-8522, Japan}
\author{Fay E. Hudson}
\affiliation{School of Electrical Engineering and Telecommunications, The University of New South Wales, Sydney, NSW 2052, Australia}
\affiliation{Diraq, Sydney, NSW 2052, Australia}
\author{Kok Wai Chan}
\affiliation{School of Electrical Engineering and Telecommunications, The University of New South Wales, Sydney, NSW 2052, Australia}
\affiliation{Diraq, Sydney, NSW 2052, Australia}
\author{Wee Han Lim}
\affiliation{School of Electrical Engineering and Telecommunications, The University of New South Wales, Sydney, NSW 2052, Australia}
\affiliation{Diraq, Sydney, NSW 2052, Australia}
\author{Andrew S. Dzurak}
\affiliation{School of Electrical Engineering and Telecommunications, The University of New South Wales, Sydney, NSW 2052, Australia}
\affiliation{Diraq, Sydney, NSW 2052, Australia}
\author{Andre Saraiva}
\email[Corresponding Author: ]{a.saraiva@unsw.edu.au}
\affiliation{School of Electrical Engineering and Telecommunications, The University of New South Wales, Sydney, NSW 2052, Australia}
\affiliation{Diraq, Sydney, NSW 2052, Australia}

\begin{abstract}
    In a full-scale quantum computer with a fault-tolerant architecture, having scalable, long-range interaction between qubits is expected to be a highly valuable resource. One promising method of achieving this is through the light-matter interaction between spins in semiconductors and photons in superconducting cavities. This paper examines the theory of both transverse and longitudinal spin-photon coupling and their applications in the silicon metal-oxide-semiconductor (SiMOS) platform. We propose a method of coupling which uses the intrinsic spin-orbit interaction arising from orbital degeneracies in SiMOS qubits. Using theoretical analysis and experimental data, we show that the strong coupling regime is achievable in the transverse scheme. We also evaluate the feasibility of a longitudinal coupling driven by an AC modulation on the qubit. These coupling methods eschew the requirement for an external micromagnet, enhancing prospects for scalability and integration into a large-scale quantum computer.
\end{abstract}

\maketitle


\textit{Introduction.}~---~Silicon-based quantum computing has great potential, combining high fidelity control \cite{xue2022quantum,mkadzik2022precision,noiri2022fast,yoneda2018quantum,huang2019fidelity,tanttu2023stability}, industrial fabrication processes \cite{ansaloni2020single,zwerver2022qubits}, and high density facilitating integration with classical electronics \cite{saraiva2022materials,burkard2023semiconductor}. Currently, two qubit gates rely on the exchange interaction, which is limited to nearest-neighbor qubits. Therefore, an important component of a full-scale quantum computer is the development of quantum interconnects or long-range coupling in a fault-tolerant architecture \cite{vandersypen2017interfacing,boter2022spiderweb,kunne2023spinbus}.

There have been several methods of long-range coupling proposed, including coherent spin transport \cite{yoneda2021coherent,seidler2022conveyor}, coherent SWAP gates \cite{sigillito2019coherent}, spin bus \cite{sanchez2014long}, surface acoustic waves \cite{sogawa2001transport}, and jellybean quantum dots \cite{wang2023jellybean,patomaki2023elongated}. In this paper, we explore the coupling of electron spin qubits in SiMOS to a superconducting cavity, which has been studied previously in Si/SiGe \cite{samkharadze2018strong,borjans2020resonant,harvey2022coherent}, GaAs \cite{landig2018coherent, Bottcher2022parametric}, and holes in SiMOS \cite{yu2023strong}. Spin-photon coupling using superconducting cavities is advantageous in that the interaction spans a much larger distance (up to millimeters) compared to other methods. However, most studies so far have required the use of a micromagnet to provide the spin-charge hybridization required for strong spin-photon coupling.

In this paper, we propose a method of spin-photon coupling which utilizes the intrinsic spin-orbit interaction observed for electrons in SiMOS devices \cite{gilbert2023demand}. Both transverse and longitudinal coupling schemes will be discussed with a view towards achieving strong spin-photon coupling. We examine a system in which fast Rabi driving was achieved and theoretically consider the addition of a qubit-resonator interaction to the system. First, we focus on transverse spin-photon coupling and estimate its expected coupling strength. Then, we move to a different regime optimized for longitudinal coupling and evaluate its viability. Finally, we discuss both the opportunities and challenges of using intrinsic spin-orbit interactions for long-range spin-photon coupling.

\begin{figure*}[ht]
     \centering
     \includegraphics[width=0.98\textwidth, angle=0, trim= 0cm 22cm 0cm 0cm]{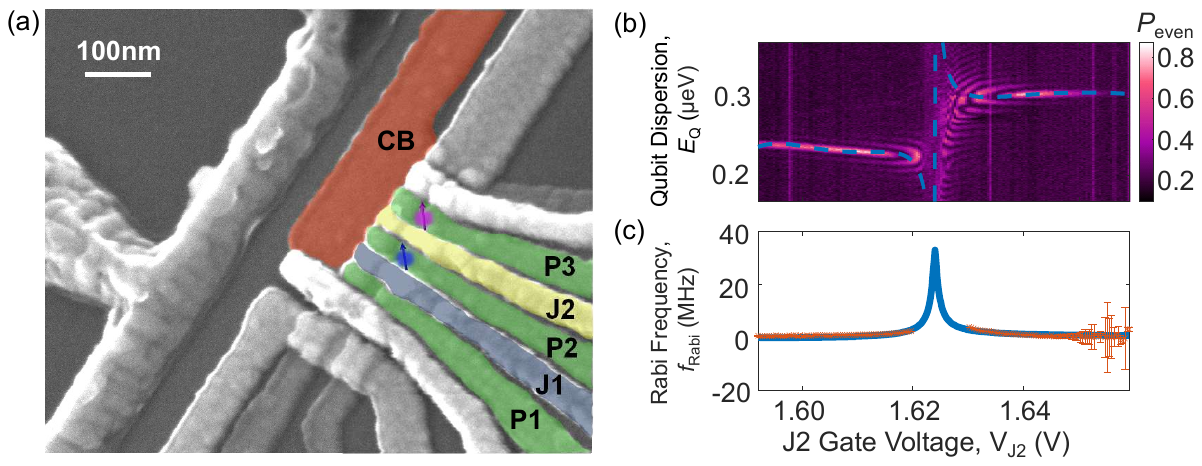}
     \caption{\label{fig:4levels} \textbf{Rabi Frequency Speed-up with EDSR.}
     (a) False-colored scanning electron microscopy (SEM) of a device nominally identical to the measured device. In the case of (b) and (c), the qubit under P3 (purple) is pulsed via a microwave pulse on the CB gate, with the qubit under P2 (blue) used as an ancilla for readout.
     (b) A PESOS map obtained with 15 dBm driving on the J2 gate in this device which was also analyzed in Ref.~\cite{gilbert2023demand} with fitted qubit dispersion (blue). Colorbar is even state probability (not re-scaled).
     (c) Extracted Rabi frequencies $f_\mathrm{Rabi}$ from the PESOS map, with a fit based on the four-level model (orange).
     }
\end{figure*}

We begin by motivating the case for strong spin-photon coupling from the perspective of electric driving in SiMOS devices. In silicon quantum dot devices, most studies of electrically driven electrons require the use of a micromagnet to achieve strong driving owing to the relatively small magnitude of spin-orbit coupling in silicon \cite{huang2017electrically,tanttu2019controlling}. However, electric driving of up to 81 MHz was observed in experiments where spin-orbit coupling effects were enhanced by operating close to an inter-orbital degeneracy point \cite{gilbert2023demand}. A schematic of the device used is shown in Fig.~\ref{fig:4levels}(a).
The voltage applied on the J2 gate is swept while generating a pulsed electron spin-orbital spectroscopy (PESOS) map [Fig.~\ref{fig:4levels}(b)], where we can observe a clear speed up of Rabi oscillations via the interference patterns that appear alongside the main resonance \cite{gilbert2023demand}. From the PESOS map, we can extract two important parameters: the qubit frequency [Fig.~\ref{fig:4levels}(b), blue], and the Rabi frequency [Fig.~\ref{fig:4levels}(c), blue]. The Rabi frequency is fitted using a four-level model which describes the quantum dot system as a single qubit with two orbital levels in the presence of magnetic field with the following Hamiltonian,
\begin{multline}
    \hat{H}_\mathrm{qd} = (E_\text{Z,A}+\eta_\text{A} V_\text{J})\frac{(\mathds{1}+\sigma_z)}{2}\otimes\sigma_z \\
    + (E_\text{Z,B}+\eta_\text{B} V_\text{J})\frac{(\mathds{1}-\sigma_z)}{2}\otimes\sigma_z \\
    + \beta_1 V_\text{J} \frac{(\mathds{1}-\sigma_z)}{2}\otimes\mathds{1} + \Delta~\sigma_x\otimes\mathds{1} \\ 
    + \Delta_\mathrm{sd}~ \sigma_x\otimes\sigma_z + \Delta_\mathrm{sf}~ \sigma_x\otimes\sigma_x
    \label{eq:Hqd}
\end{multline}
where $E_\text{Z,A(B)}$ are the Zeeman splittings for each of the orbitals, $\eta_\text{A(B)}$ are the respective linear Stark shifts, $\beta_1$ is the relative lever arm between the two orbitals, $V_\text{J}$ is the applied gate voltage on the J gate (either J1 or J2 depending on the regime), $\Delta$ is the coupling between the orbitals, $\Delta_\text{sd}$ is the spin-dependent component of spin-orbit coupling, and $\Delta_\text{sf}$ is the spin-flip component of the spin-orbit coupling \cite{feng2022control}. To obtain the Rabi frequency from this model, we consider the Hamiltonian with driving amplitude $\Omega_\text{AC}$,
\begin{align}
    \hat{H}_\text{AC} = \Omega_\text{AC}~ \sigma_x\otimes\sigma_z
\label{eq:AC}
\end{align}
With this, the Rabi frequency is defined as the coupling between the ground, $\ket{g}$, and the 1st excitation state, $\ket{e}$, of $\hat{H}_\text{AC}$ in the eigenbasis of Eq.~(\ref{eq:Hqd}), and can be calculated as $f_\text{Rabi}=\left|\bra{e}\hat{H}_\text{AC}\ket{g}\right|$. The fitting protocol follows that which was used in Ref.~\cite{gilbert2023demand}, and we provide the fitted parameters for all relevant regimes in the Supplementary material \cite{supp}.

So far, the fast electrically driven spin resonance (EDSR) is obtained by driving at high microwave powers, which is not the case when the spins are coupled via a superconducting microwave resonator in the single-photon regime. To that end, we adopt a model of a qubit-resonator system from which coupling strengths can be estimated by assuming similar resonator parameters to previous experiments \cite{borjans2020resonant,benito2019optimized,corrigan2022longitudinal}.
The total qubit-resonator system can be described by the following Hamiltonian:
\begin{align}
    H_\text{tot} = H_\text{res} + H_\text{qd} + H_\text{int}
\end{align}
where $H_\text{qd}$ has been defined above, and $H_\text{res}=\hbar\omega_\text{res}a^\dagger a$. The interaction between the qubit and the resonator is given by $H_\text{int}$ and can be written in the following form \cite{ruskov2019quantum,ruskov2021modulated},
\begin{multline}
    H_\text{int} = \hbar [g_\perp\sigma_x + g_\parallel\sigma_z\cos(\omega_m t + \varphi_m)] (a + a^\dagger) \\+ \hbar \delta\omega \sigma_z (a^\dagger a + \frac{1}{2})
    \label{eq:Hint}
\end{multline}
Here, $g_\perp$ is the transverse component of the qubit-resonator coupling. As we will show, applying a sinusoidal modulation on the qubit with frequency $\omega_m$ and phase $\phi_m$ gives a dynamical longitudinal coupling, denoted by $g_\parallel$. There is also an energy-energy dispersive coupling $\delta\omega$. Both transverse and longitudinal schemes will be explored in this paper.

\begin{figure*}
      \centering
      \includegraphics[width=0.98\textwidth, angle = 0, trim = 0cm 18cm 0cm 0cm]{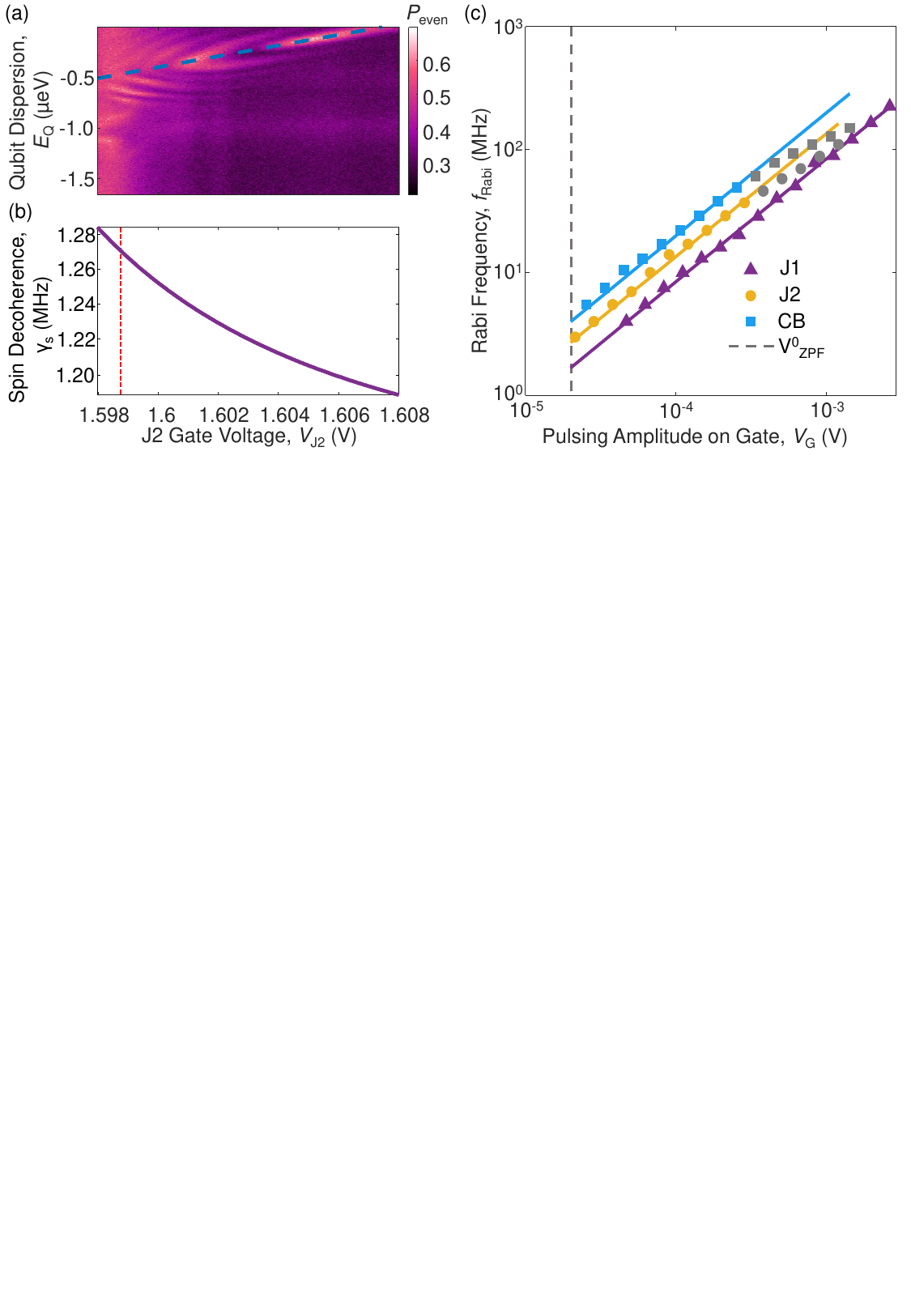}
      \caption{\label{fig:transverse} \textbf{Rabi Frequency at Single-Photon Level.}
      (a) PESOS map taken at $B_0=700~\text{mT}$, in the same electronic configuration as Fig.~\ref{fig:4levels}(b), at 20 dBm driving on the CB gate. Blue dashed line shows the fitted qubit dispersion.
      (b) Spin decoherence as calculated in \cite{supp}. Red dashed line indicates J2 gate voltage at which we measure the Rabi frequencies, with the corresponding $\gamma_\mathrm{s}=1.27~\mathrm{MHz}$.
      (c) Rabi frequency, $f_\mathrm{Rabi}$, plotted against microwave power applied on the CB (blue), J1 (yellow), and J2 (purple) gates. A linear fit is used to extrapolate to the single-photon level (20 \textmu{}V indicated by the black dashed line). The extrapolated Rabi frequency are respectively 3.98 MHz (CB), 1.68 MHz (J1), and 2.70 MHz (J2). Some data points for the CB and J2 gate fits are excluded from the fit due to a tapering effect observable at higher powers (grey).
      }
\end{figure*}

\textit{Transverse coupling.}~---~To estimate the expected transverse spin-photon coupling strength in the SiMOS platform, we experiment on a device without a resonator [Fig.~\ref{fig:4levels}(a)] and extrapolate the observed Rabi frequencies to the expected single-photon voltage fluctuations $V^0_\mathrm{ZPF}$ of a high-impedance resonator. In the eigenbasis of $H_\mathrm{qd}$, the resonator provides an electric drive equivalent to Eq.~(\ref{eq:AC}) with drive amplitude
\begin{align}
   \Omega_\text{AC,ZPF} = \frac{\alpha_c}{2}V^0_\mathrm{ZPF}
\end{align}
where $\alpha_c$ is lever arm of the resonator gate.
The spin-photon coupling is
\begin{align}
    g_\perp/2\pi = f_\mathrm{Rabi,ZPF}
\end{align}

Fig.~\ref{fig:transverse}(a) shows the PESOS map of the device in an external magnetic field of 700 mT, where we operate close to the orbital degeneracy point at $V_\mathrm{J2} = 1.598~\mathrm{V}$. In this regime there is a speedup in the Rabi frequency due to the increased spin-orbit coupling of the electron \cite{gilbert2023demand}. While the Zeeman splitting induced by the magnetic field is higher than the expected cavity resonant frequency ($E_z = 19~\mathrm{GHz}, \omega_\mathrm{r}/2\pi = 6~\mathrm{GHz}$), the Rabi speedup near the degeneracy is not expected to differ significantly for lower magnetic fields at which the qubit is on resonance with the cavity ($B_0 \approx 300~\mathrm{mT}$), since the speedup is dependent only on the electric field strength. 

Rabi oscillations were measured at a voltage point close to the degeneracy with a microwave drive applied to three gates: CB, J1, and J2 [Fig.~\ref{fig:4levels}(a)]. Rabi frequencies were measured as a function of the microwave power applied to the corresponding gate. We then convert the microwave power into voltage applied at the gate, which takes into account the difference in lever arm between gates \cite{supp}. As the drive voltage decreases, $f_\mathrm{Rabi}$ drops linearly, which can be observed in Fig.~\ref{fig:transverse}(c).  We observe a tapering effect for voltage amplitudes above $0.7~\mathrm{mV}$, [points in grey, Fig.~\ref{fig:transverse}(c)]. This behavior has been previously observed in previous experiments on similar devices \cite{gilbert2023demand,vahapoglu2022coherent}, and we exclude these data points from the fit since this effect is not relevant at single-photon levels.

\begin{table}[ht]
    \begin{center}
    \begin{tabular}{c|c|c|c|c} 
        Driven Gate \; & $g_s/2\pi$ & $\kappa_r/2\pi$ & $\gamma_s/2\pi$ & $\frac{2g_s}{\gamma_s+\kappa_r/2}$ \\
        \hline
        CB & \; 3.98 MHz \; & \; 2 MHz \; & 1.27 MHz & 3.5 \\ 
        J1 & \; 1.68 MHz \; & \; 2 MHz \; & 1.27 MHz & 1.5 \\
        J2 & \; 2.70 MHz \; & \; 2 MHz \; & 1.27 MHz & 2.4\\
    \end{tabular}
    \caption{Table of spin-photon coupling parameters for each of the three driving gates.}
    \label{tab_fit_values}
    \end{center}
\end{table}

We then estimate the voltage amplitude at single-photon levels for a typical high impedance resonator ($\omega_\mathrm{r} = 6~\mathrm{GHz}, Z_\mathrm{r} = 2~\mathrm{k\Omega}$) \cite{samkharadze2016high,holman20213d, yu2021magnetic}. The zero point voltage fluctuation (amplitude) of a single photon is 
\begin{align}
    V^0_\text{ZPF} = \omega_r \sqrt{\frac{2Z_r}{\pi\hbar}}
\end{align}
giving a value of approximately $20~\text{\textmu{}V}$. Extrapolating the fits in Fig.~\ref{fig:transverse}(c) to the single photon level, we obtain Rabi frequencies of $3.98~\mathrm{MHz}$, $1.68~\mathrm{MHz}$, and $2.70~\mathrm{MHz}$ when driving on the CB, J1, and J2 gates respectively. 

The decoherence in the system is calculated from the first and second derivatives of the fitted qubit dispersion \cite{Russ2016coupling,supp} and is plotted in Fig.~\ref{fig:transverse}(b). We assume that the dominant source of decoherence for a qubit is charge noise, as previously observed for spin qubits in isotopically enriched silicon \cite{yoneda2021coherent,feng2022control,chan2018assessment}. Measurements for all three gates were taken at the same voltage relative to the degeneracy, so the decoherence is expected to be the same for all driving gates.

In order to evaluate the coupling strength of our system, we use the criterion $2g_s > \gamma_s+\kappa_r/2$ to determine if the strong coupling regime is achieved. When this criteria is satisfied, the vacuum Rabi splitting $2g_s$ is larger than the combined qubit and resonator linewidths, and a splitting of the resonator’s frequency response is observable \cite{blais2021circuit}. We assume the resonator has a decay rate $\kappa_\mathrm{r}/2\pi$ of $2~\mathrm{MHz}$, comparable to previous experiments with high impedance resonators on Silicon substrates \cite{HarveyCollard2022coherent,corrigan2022longitudinal,borjans2020resonant}. With this, we calculate the expected coupling strength and find that the strong coupling regime is achievable for all three driving gates (Tab.~\ref{tab_fit_values}). The strongest coupling is obtained with the CB gate, due to its high lever arm caused by the gate geometry and proximity to the electron \cite{borjans2020split}.

We note that the transverse coupling scheme can be vulnerable to spin relaxation due to the Purcell effect \cite{bienfait2016controlling}, which could be mitigated by detuning the qubit from the resonator \cite{HarveyCollard2022coherent}.

\begin{figure}[ht]
      \centering
      \includegraphics[width = 0.95\columnwidth, angle = 0, trim = 0cm 17.5cm 9.6cm 0cm]{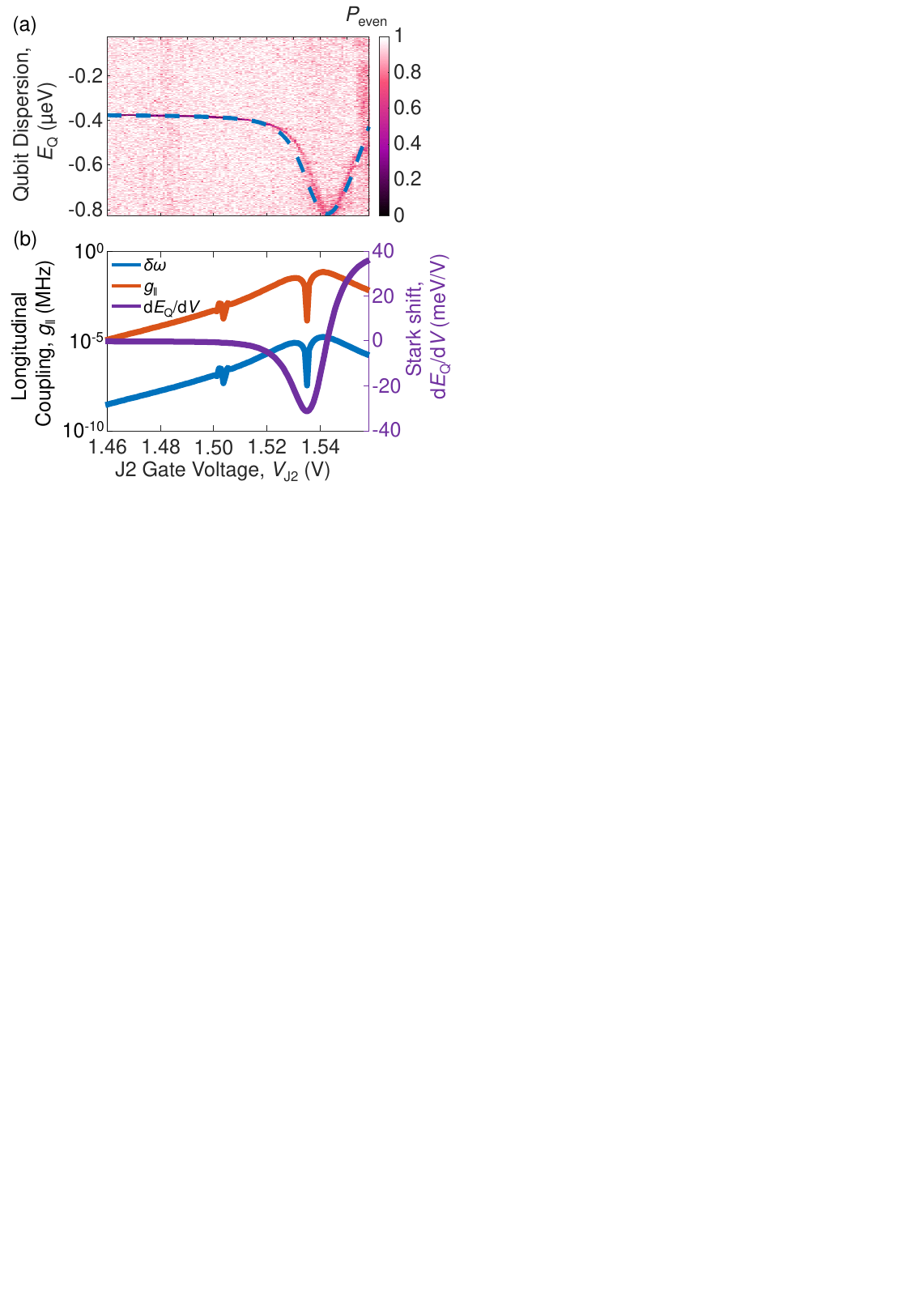}
      \caption{\label{fig:longitudinal} \textbf{Longitudinal Coupling Strengths.}
      (a) PESOS map taken at $B_0=620~\text{mT}$.
      (b) Magnitudes of the different longitudinal couplings (dispersive and dynamical longitudinal couplings) for the map in (a). The second order Stark shift is plotted in purple.
      }
\end{figure}

\textit{Longitudinal coupling.}~---~We now focus our attention on a different spin-photon coupling mechanism: the longitudinal coupling. In this scheme, we are no longer restricted to be on resonance with the cavity, and can make use of a sweet spot in detuning, where coherence times are enhanced.
At the sweet spot, the first order static coupling term disappears since the first derivative of the qubit dispersion goes to zero ($\partial E_\text{Q}/\partial V_\text{G}=0$). However, there remains an energy-energy coupling because the second derivative at that point is non zero, $\partial^2E_\mathrm{Q}/\partial V_\text{G}^2 \neq 0$, which leads to a non-zero coupling strength $\delta\omega$, given by
\begin{align}
    \delta\omega = \frac{\alpha_\text{c}^2}{2} \left(V^0_\text{ZPF}\right)^2 \frac{\partial^2 E_\text{Q}}{\partial V^2}
    \label{eq:omega}
\end{align}

The $\delta\omega$ term is likely too small for a viable spin-photon coupling scheme. However, the energy-energy coupling can be enhanced by applying an AC modulation to the gate voltage of our qubit \cite{ruskov2021modulated, harvey2018coupling}. 
By choosing a modulation frequency equal to the resonator's frequency $\omega_r$, we can introduce an additional term to the interaction Hamiltonian containing $g_\parallel$ in Eq.~(\ref{eq:Hint}) with the coupling strength \cite{ruskov2021modulated},
\begin{align}
    g_\parallel = \frac{\alpha_\text{c}}{2} V^0_\text{ZPF} \frac{\partial^2 E_\text{Q}}{\partial V^2} \tilde{V}_\text{m}
    \label{eq:gdy} 
\end{align}
where $\tilde{V}_\text{m}$ is the modulation amplitude on the qubit. Since we are driving on resonance with the cavity frequency, there exist other terms that will be canceled out in the rotating wave approximation \cite{ruskov2021modulated,corrigan2022longitudinal} which are omitted in our calculations here. We assume a modulation amplitude of about $20~\mathrm{mV}$, consistent with the maximum applicable amplitude in electric driving experiments \cite{gilbert2023demand}.%

Fig.~\ref{fig:longitudinal}(a) is the PESOS map taken from the same device as in Fig.~\ref{fig:4levels}, but at a different electronic configuration and magnetic field($B_0=620~\mathrm{mT}$). Here, the qubit being pulsed is under the P1 gate [labeled in Fig.~\ref{fig:4levels}(a)] with the qubit under P2 gate used as the ancilla for measurement. We examine this regime specifically because there is a clear sweet spot ($\partial E_\text{Q}/\partial V=0$) in the qubit dispersion. There is also a sizable second order Stark shift which is conducive to achieving dispersive ($\delta\omega$) coupling. We calculate the longitudinal couplings using Eqs.~(\ref{eq:omega}) and (\ref{eq:gdy}), which are plotted in Fig.~\ref{fig:longitudinal}(b) alongside the Stark shift, $\partial E_\mathrm{Q}/\partial V$, We obtain a peak value for $g_\parallel=69~\mathrm{kHz}$ at $1.54~\mathrm{V}$.

We now estimate the tunability ratio $g_\parallel/2\delta\omega$ as a metric of how strong the longitudinal coupling is compared to the always-on dispersive coupling at a fixed operation point \cite{corrigan2022longitudinal}. In the voltage configuration considered in Fig.~\ref{fig:longitudinal}(a), we are able to achieve tunability ratios of about 4000, which is larger than previously reported for electrons in silicon \cite{corrigan2022longitudinal}. This is likely due to the higher gate lever arms of the SiMOS platform and the larger tunnel coupling of the qubit, allowing for a higher modulation amplitude. The coupling can also be turned on and off by moving the qubit in voltage space away from the ideal operation point with maximal coupling.

In the longitudinal coupling scheme, the qubit is not required to be on resonance with the cavity and the device can be operated in the adiabatic regime where $\hbar \omega_r \ll E_Q$. This reduces cavity-induced relaxation via the Purcell effect \cite{corrigan2022longitudinal} and allows for greater flexibility when tuning the qubit or external magnetic field.

In our proposal for spin-photon coupling, we rely significantly on our ability to find non-linear Stark shift in our qubits which in our studies of electrical control have been shown to be highly reproducible \cite{gilbert2023demand}. There are several ways we can tune our qubit dispersion to improve the longitudinal coupling. We can electrically tune the orbital coupling term $\Delta$ using neighboring gates \cite{gilbert2023demand}. Varying the magnitude and angle of the external magnetic field is also beneficial \cite{tanttu2019controlling} but with the caveat that this is of limited use when scaling to larger qubit systems.

\textit{Conclusion.}~---~Based on experimental studies of the speed-up in Rabi frequencies using electrical drive in our devices, we have identified voltage regimes where strong transverse spin-photon coupling can be obtained given a suitable superconducting resonator. In addition, we also find voltage regimes where we can operate with significant longitudinal coupling, which removes the need to match the qubit frequency to the resonator frequency. These methods use an intrinsic spin-orbit coupling and do not require a micromagnet, reducing fabrication and engineering constraints when scaling beyond two qubits. Further improvements to the spin-photon coupling strength could be made by reducing qubit charge noise through fabrication \cite{elsayed2022low} and increasing the lever arm with optimized gate design \cite{borjans2020split}. The resonator decay rate could also be improved with better gate filtering \cite{harveycollard2020onchip}, or by moving the resonator off-chip \cite{corrigan2022longitudinal}. We believe that these studies show that there is potential to achieve strong spin-photon coupling in SiMOS devices, paving the way for long-range coupling in scalable architectures for silicon quantum computing.

\begin{acknowledgments}
    We thank A. Dickie and S. Serrano for hardware support with the experiments. We thank B. Harpt and R. de Melo e Souza for helpful discussions. We acknowledge support from the Australian Research Council (FL190100167, CE170100012, and IM230100396) and the US Army Research Office (W911NF-23-10092). The views and conclusions contained in this document are those of the authors and should not be interpreted as representing the official policies, either expressed or implied, of the Army Research Office or the US Government. M.K.F. and J.Y.H. acknowledge support from the Sydney Quantum Academy.
\end{acknowledgments}

\bibliographystyle{apsrev4-2}
\bibliography{sp_refs}

\pagebreak
\widetext
\begin{center}
\textbf{\large Supplementary Material: Methods for transverse and longitudinal spin-photon coupling in silicon quantum dots with intrinsic spin-orbit effect}
\end{center}
\setcounter{equation}{0}
\setcounter{figure}{0}
\setcounter{table}{0}
\setcounter{page}{1}
\makeatletter
\renewcommand{\theequation}{S\arabic{equation}}
\renewcommand{\thefigure}{S\arabic{figure}}
\renewcommand{\thetable}{S\arabic{table}}
\renewcommand{\bibnumfmt}[1]{[S#1]}
\renewcommand{\citenumfont}[1]{S#1}
\vspace{0.8cm}

\section{Microwave Power Estimation}

\begin{figure}[ht]
      \centering
      \includegraphics[width = 0.5\textwidth, angle = 0, trim = 0cm 22cm 12.5cm 0.6cm]{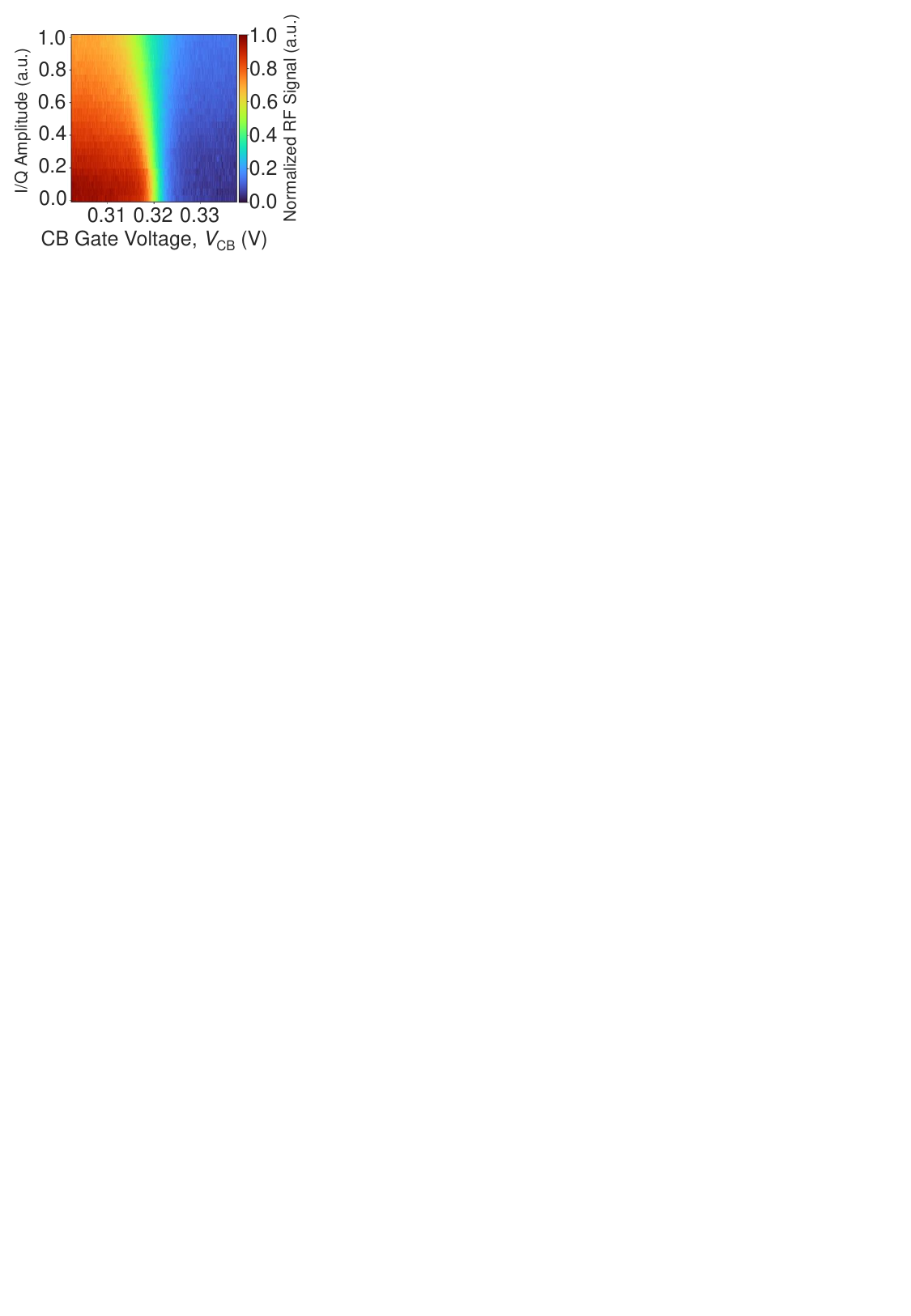}
      \caption{\label{suppfig:mw}
      \textbf{Microwave broadening on the CB gate.} SET signal as a function of IQ amplitude and CB gate voltage around the interdot transition, $V_\mathrm{CB}$.
      }
\end{figure}

In order to understand the amount of power being delivered to the gate, we measured the broadening of the interdot transition as a function of CB gate voltage, $V_\mathrm{CB}$, and the IQ amplitude. (Fig.~\ref{suppfig:mw}). We observe a clear broadening of the transition at higher microwave powers, which we assume follows a Fermi-Dirac distribution. While this technique does not account for heating effects at higher powers, any thermal broadening would be additive. Therefore, attributing all broadening to the microwave source gives an upper bound for the driving amplitude and a lower bound for the Rabi frequency at the single photon level. With this, we estimate a minimum line attenuation of 77dB, 75dB, and 71dB from the microwave source to the CB, J2, and J1 gates respectively. This calculation is used to obtain the results in  Fig.~\ref{fig:transverse}(b).

\section{Rabi Oscillations}

\begin{figure}[ht]
      \centering
      \includegraphics[width = 0.9\textwidth, angle = 0, trim = 0cm 20.5cm 0cm 0cm]{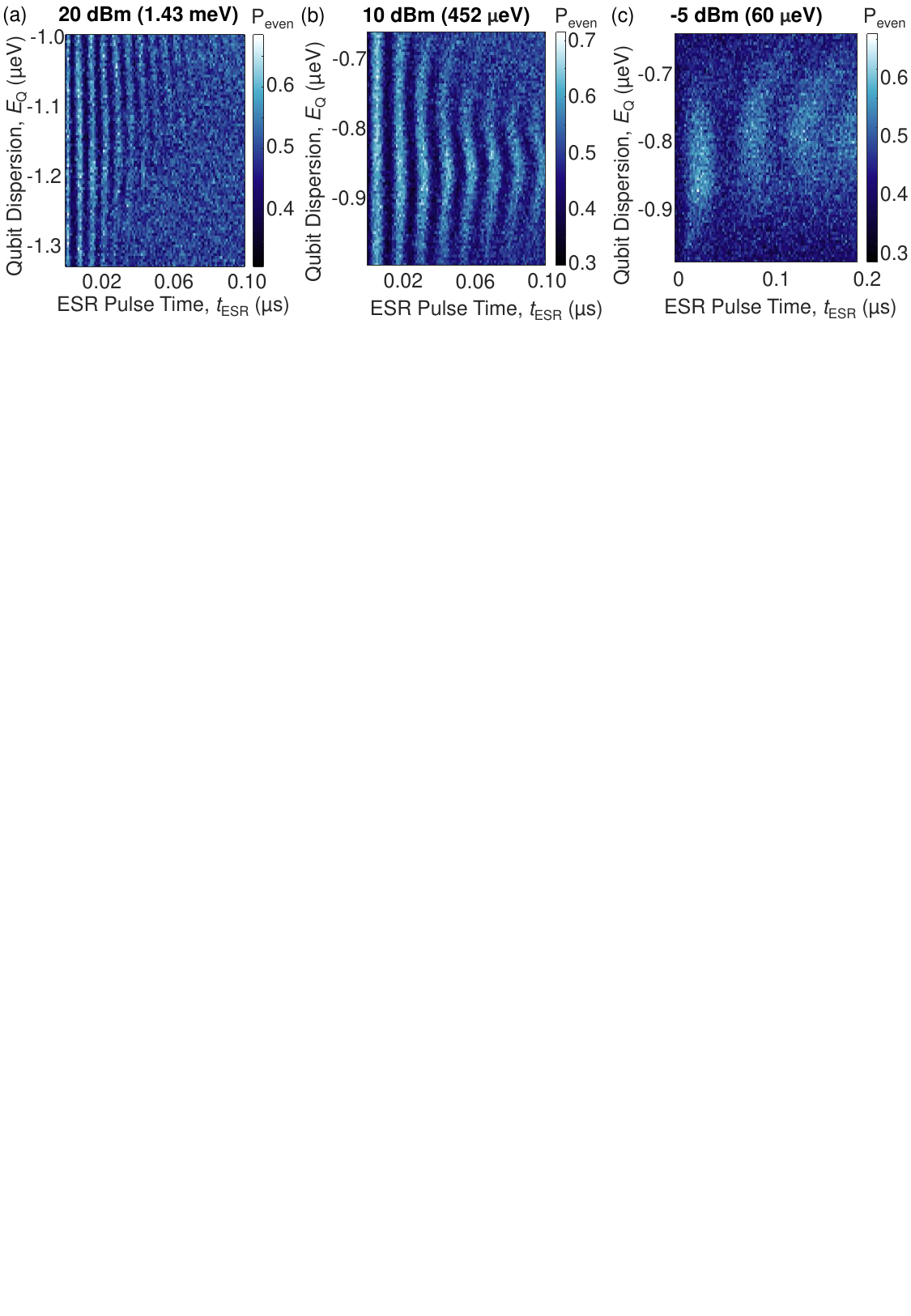}
      \caption{\label{suppfig:rabi} \textbf{Rabi oscillations at different powers.}
      (a) Rabi oscillations measured at 20 dBm (equivalent to 1.4 mV AC voltage amplitude on the CB gate), 
      (b) Rabi oscillations measured at 10 dBm (450 \textmu{}V), 
      (c) Rabi oscillations measured at -5 dBm (60 \textmu{}V).
      }
\end{figure}

In Fig.~\ref{fig:transverse}(b), we extrapolated the Rabi frequencies as a function of the amplitude of the driven voltage for driving on CB, J1, and J2 gates. Here, we show selected Rabi oscillations that were obtained by driving on CB for different powers (Fig.~\ref{suppfig:rabi}). We demonstrate here the trend in the Rabi frequency with decreasing powers and also include off-resonant oscillations to show that they are independent of the qubit dispersion.

By taking measurements with different microwave powers and driving gates, we can extract fitted Rabi frequencies, corresponding to the results in Fig.~\ref{fig:transverse}(b) in the main text.

\section{Fitting parameters}

\begin{center}
\begin{table}[h!]
\resizebox{0.8\textwidth}{!}{%
\begin{tabular}{ c c c c c c }
\hline\hline
\textbf{Electron Configuration} & \textbf{A} & \textbf{B}  & \textbf{C}\\ 
\hline
$B_0$ (mT) & $700$ & $700$ & $620$ \\
$V_0$ (V) & $1.596\pm0.002$ & $1.581\pm0.5$ & $1.537\pm0.002$ \\
$\alpha_\mathrm{rel}$ (THz/V) & 0.5 & 0.5 & 0.5 \\
$\Delta$ (GHz) & $6.6\pm0.6$ & $0.113\pm800$ & $2.5\pm0.7$ \\
$\Delta_\mathrm{sd}$ (GHz) & $-0.15\pm0.02$ & $-0.304\pm1000$ & $0.02\pm0.03$ \\
$\Delta_\mathrm{sf}$ (GHz) & $-0.163\pm0.007$ & $-0.061\pm200$ & $-0.002\pm0.4$ \\
$dE_\mathrm{Z}$ (GHz) & $-0.327\pm0.07$ & $0.996\pm2000$ & $-0.25\pm0.08$ \\
$\eta_\mathrm{A}$ (MHz/V) & $-6582\pm2000$ & $-6000\pm10^5$ & $16\pm60$ \\
$\eta_\mathrm{B}$ (MHz/V) & $429\pm90$ & $10^5\pm2\times10^7$ & $-5003\pm2000$ \\
$c$ (MHz) & $195\pm20$ & $-300\pm9000$ & $-340\pm100$ \\
$\Omega_\mathrm{AC}$ (MHz) & $65\pm5$ & $0.04\pm100$ & Not Fitted \\
\hline\hline
\end{tabular}}
\caption{Fitted parameters of the four-level model for the different regimes. Configuration A refers to the regime in Fig.~\ref{fig:4levels}, configuration B refers to the regime in Fig.~\ref{fig:transverse}, and configuration C refers to the regime in Fig.~\ref{fig:longitudinal}. Configuration B has extremely large error bars indicating that these fitted parameters cannot be trusted to describe the full system, but is valid only within the experimental range.}
\label{table:fittedvals}
\end{table}
\end{center}

In all of the regimes discussed in the main text, we used a four-level model including orbital and spin states to fit the qubit dispersion and obtain the energy spectrum of the qubit \cite{gilbert2023demand}. We acknowledge that the fitting for configuration B works only within the fitted voltage range, as indicated in Fig.~\ref{fig:transverse}. Table \ref{table:fittedvals} lists the fitted parameters.

\section{PESOS map and Energy diagram at 0.7 T}

\begin{figure}[ht]
      \centering
      \includegraphics[width = 0.7\textwidth, angle = 0, trim = 0cm 21.5cm 6.4cm 0cm]{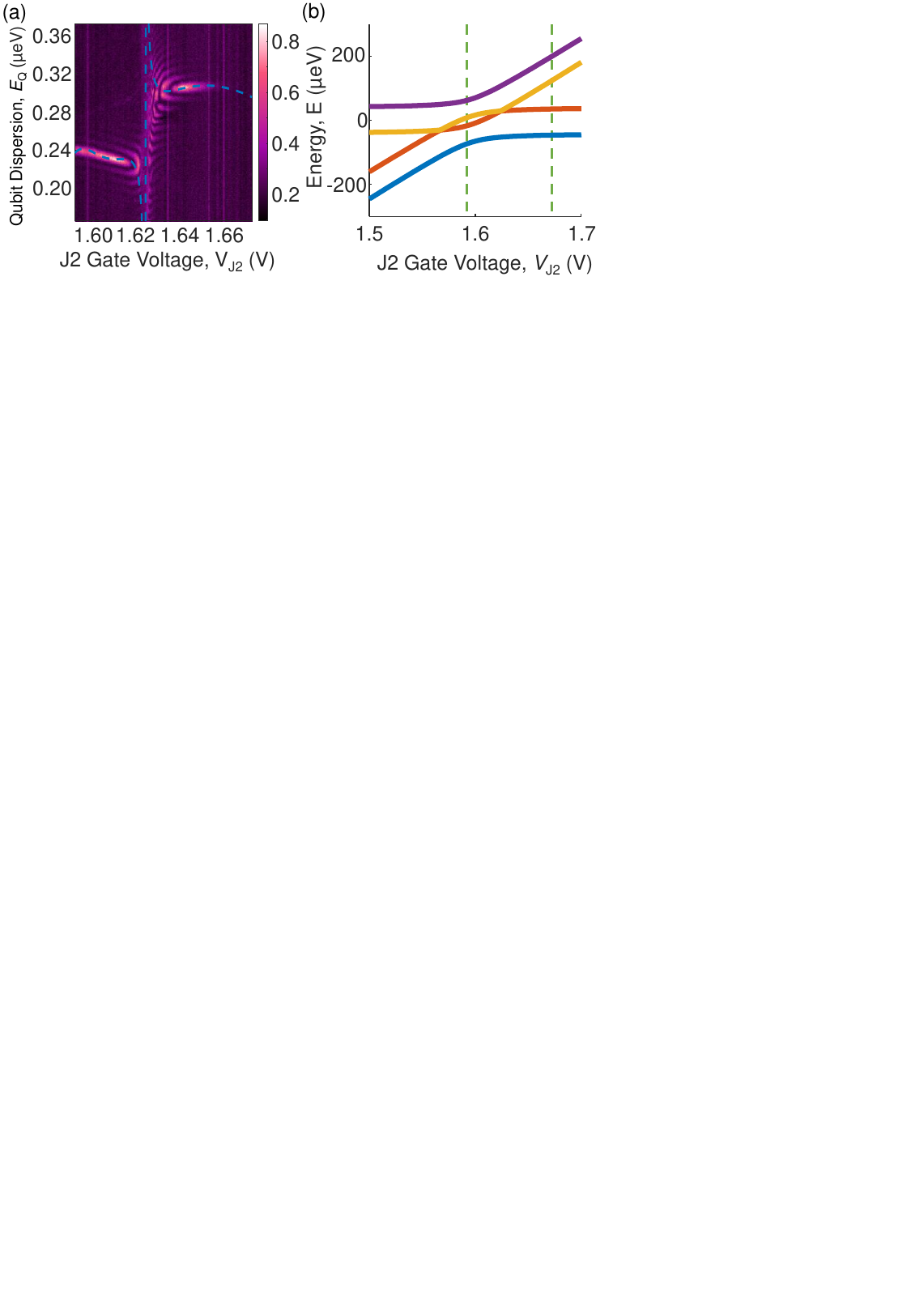}
      \caption{\label{suppfig:700mT} \textbf{Energy Diagram and Longitudinal Coupling at 700 mT.}
      (a) The full version of the PESOS map shown in Fig.~\ref{fig:4levels}(b). The fitting beyond 1.65 V is unreliable because a resonance can no longer be observed.
      (b) Energy diagram of the system given the fitted values. The green dashed lines mark the region where the qubit dispersion data was fitted.
      }
\end{figure}

In the main text, we explored the operation of transverse couplings in the 700 mT regime and here, we show the full PESOS map, including regions where we excluded from the fit (high voltage region in Fig.~\ref{suppfig:700mT}(a)). We also show here the full energy diagram plotted using the fitted values with the green dashed lines marking out the area where the fit was performed.

\section{PESOS map and Energy diagram at 0.62 T}

\begin{figure}[ht]
      \centering
      \includegraphics[width = 0.7\textwidth, angle = 0, trim = 0cm 20.5cm 6.2cm 0cm]{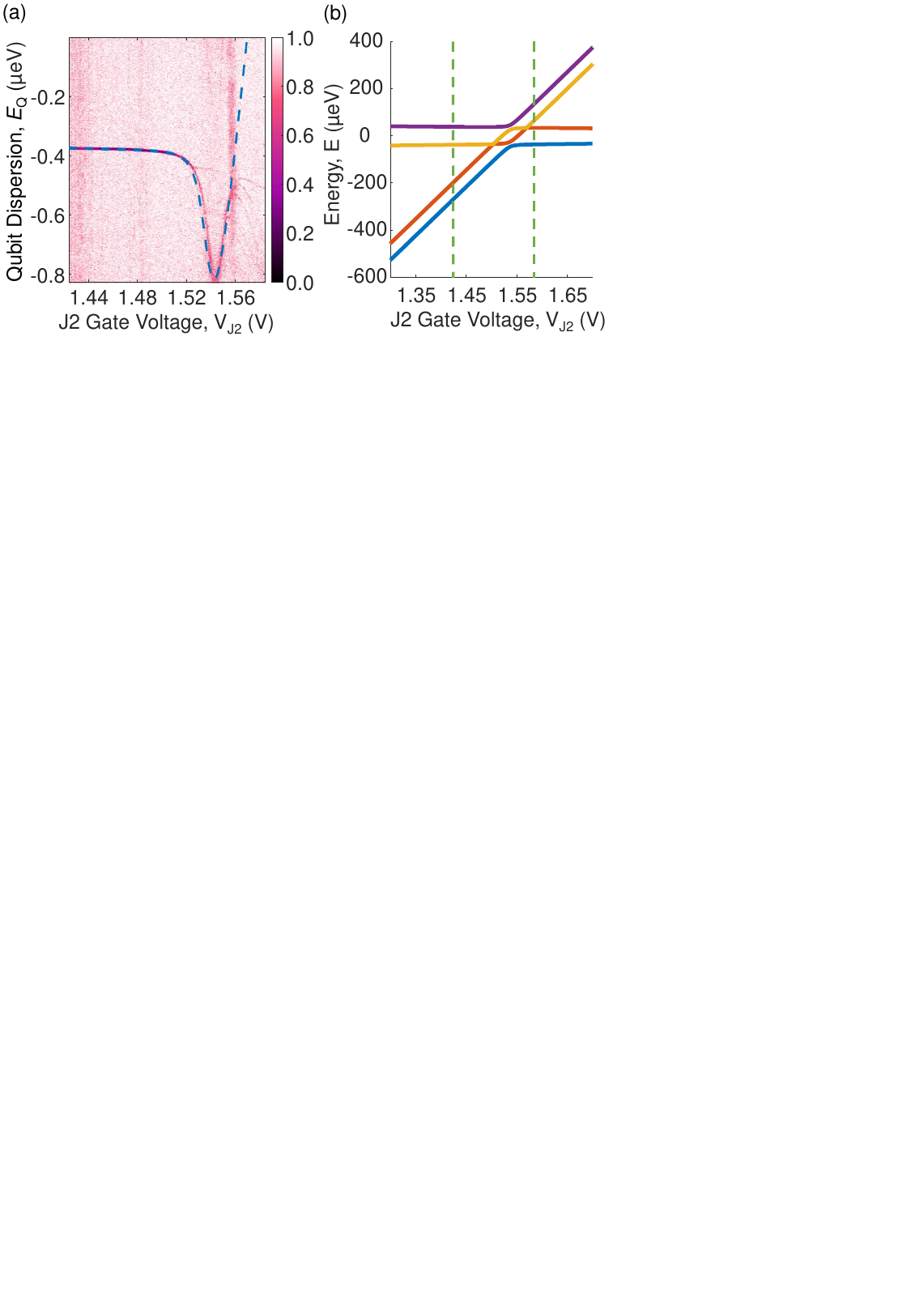}
      \caption{\label{suppfig:620mT} \textbf{Energy diagram and cooperativity at 620 mT.}
      (a)  PESOS map measured at 620 mT. Beyond 1.56 V, the fit is unreliable due to the appearance of multiple resonances.
      (b) Energy diagram of the system given the fitted values. The green dashed lines mark the region where the qubit dispersion data was fitted.
      }
\end{figure}

In the main text, we explored longitudinal coupling at 620 mT. Here we show the full voltage range of the obtained PESOS map (Fig.~\ref{suppfig:620mT}(a)), along with the fitted energy diagram (Fig.~\ref{suppfig:620mT}(b)). The green dashed lines mark the voltage range where the fitting was performed, which is confined to only one side of the degeneracy.

\section{Calculation of Decoherence}
\label{supp:decoherence}
The expected spin decoherence in our system is calculated using the following relation \cite{Russ2016coupling},
\begin{align}
   \frac{\gamma_s}{2\pi} =\frac{1}{h} \sqrt{\frac{1}{2} \left( \frac{\partial E_\text{Q}}{\partial V} \right) ^2 \delta V_\text{noise}^2 + \frac{1}{4} \left(\frac{\partial^2 E_\text{Q}}{\partial V^2}\right)^2\delta V_\text{noise}^4}
\end{align}
where $\delta V_\mathrm{noise}$ is the voltage noise fluctuation of the gate due to charge noise, assumed to follow a $1/f$ distribution. This expression accounts for noise up to the second order, such that the coherence does not go to infinity at the theoretical sweet spot where $\partial E_\mathrm{Q}/\partial V = 0$. We estimate $\delta V$ to be 2 \textmu{}eV at 1 Hz based on results obtained in similar experimental setups \cite{tanttu2023stability}. This allows us to calculate the decoherence shown in Fig.~\ref{fig:transverse}.

\end{document}